\documentclass{ifacconf}

\usepackage{graphicx}      % include this line if your document contains figures
\usepackage{natbib}        % required for bibliography
\usepackage{psfrag}
\usepackage{amsmath} % assumes amsmath package installed
\usepackage{amssymb}  % assumes amsmath package installed
\usepackage[table]{xcolor}
\begin{document}
\begin{frontmatter}

% 21 Nov. 2025: Acknowledgement of Submission 1174 for IFAC WC 2026
%You successfully uploaded the manuscript file IFAC_WC_26___PID_achitectures (2).pdf
%Submission 1174 was successfully entered into the database
%An acknowledgement was sent to skoge@ntnu.no

\title{Advanced PID architectures for tracking changing active constraints}% in complex processes}%\thanksref{footnoteinfo}} 

%\thanks[footnoteinfo]{This project has been partially funded by...}

\author[NTNU]{Sigurd Skogestad} 
%\author[UAL]{Manuel Berenguel}
\address[NTNU]{Department of Chemical Engineering, Norwegian University of Science and Technology (NTNU), Trondheim (e-mail: sigurd.skogestad@ntnu.no)
\\ {\bf Extended version (with simulations) of paper at IFAC World Congress, August 2026, Korea. This version (with also separator case study I): \today}
}

\begin{abstract}                % 50-100 WORDS
Advanced regulatory control (ARC), also known as advanced PID architectures, is a simple and robust way of controlling processes with changing and possibly conflicting constraints, where it previously was believed - at least in academia - that model-based solutions, such as MPC, were the only effective solution. To illustrate this, ARC is applied in two case studies. The first is a gas-liquid separation process, in which selectors and split-parallel control are combined to achieve bidirectional inventory control in which the throughput manipulator moves automatically to the most optimal position. The second case study is on keeping acceptable air quality (CO2-level) and temperature in a room (in this case, a barn for cows). The CO2 and temperature constraints can be conflicting, leading to a hierarchical switching network of PID controllers.

\end{abstract}


\end{frontmatter}
%===============================================================================

\section{Introduction}

An active constraint is a constraint which, in the current situation, should be kept at its limiting (maximum or minimum) value to achieve optimal (economic) operation at steady state. The ``current situation'' is defined by the current value of disturbances, parameters, and prices in the cost function.
To achieve optimal economic operation, the most important is usually tracking and controlling active constraints \citep{Maarleveld1970}. 

For about 40 years, since the 1980s, there has been a myth, at least in the academic community, that this requires model-based schemes, such as model predictive control (MPC) and real-time optimization (RTO). 
Of course, academic researchers had some knowledge about industrial schemes for dealing with constraints, including selectors and split-range control, but 
such ``override'' schemes (e.g., as described in \cite{Maarleveld1970} for an industrial distillation column) were thought to be {\em ad hoc}, old-fashioned and not optimal, and not worth the effort of more detailed studies. Until about 10 years ago this was also my view.

{\bf Selectors.} Consider the case where a single manipulated variable (MV) (input) should control more than one controlled variable (CV) (output). MAX- and MIN-selectors are simple logic elements (sometimes referred to as overrides) for switching the CV (output) in response to a change in the active output constraints. Only one CV should be controlled at a time, which is known as {\em CV-CV switching}. 
%MAX-selectors are used for constraints that are satisfied by a large input, whereas MIN-selectors are used for constraints that are satisfied by a small input. If all constraints are satisfied by a large input, then only a single MAX-selector is needed. Then, in a given situation, one output (CV) is controlled at its (active) constraint, whereas the other constraints are over-satisfied. However, if there is also a constraint that is satisfied by a small input, requiring a MIN-selector, then there may be situations (disturbances) when there is no feasible solution (input) because the constraints are conflicting. In this case, one needs to decide which constraint has highest priority and put the corresponding selector (MIN or MAX) at the end of the decision sequence (closest to the MV) as illustrated later for case study IIB (Figure~\ref{fig:COW3}).

{\bf Split-range and split-parallel control}. In this paper, we also consider the case where two or more MVs are used to control a single CV. Only one MV should be used at a time, which is known as {\em MV-MV switching}. The traditional solution is to use split-range control, but this requires the user to specify the MV values (say, 0\% and 100\% in the simple case of valve saturation) where switching should occur. This is avoided with split-parallel control where each MV has its own controller with a different CV setpoint. Here, the switching occurs by feedback when an MV saturates (or is taken over by another controller) and control of the CV is lost. 

These and other ARC schemes were discussed in a recent paper \citep{Skogestad2023}. 
The objective of the present paper is to extend these findings (with some new rules) and show with two case studies, 
%including case study I on bidirectional inventory control, 
the power of selectors and split-parallel control. 
%The two schemes may be combined to achieve MV-CV switching, as shown for bidirectional inventory control in case study I.

%The paper is organized as follows. Section 2 presents case study I

\section{Case study I: %Inventory control and moving TPM for 
Gas-liquid separator}

\subsection{General about inventory control and TPM} 

Maintaining throughput and managing and controlling inventories (e.g., levels) are of vital importance for process operation. An important concept is the throughput manipulator (TPM), which is the variable that sets the throughput (production rate) for the entire process. Most processes have only one TPM, which also sets the 
feed rate(s) and product rate(s). 
Additional feeds are usually set by ratio control to the main feed.
%For cases with multiple feeds, the feeds are usually set in ratio to each other for operational reasons. 
The location of the TPM has a very strong impact on the resulting control system. The TPM is traditionally located at the (main) feed or at the (main) product, but in general it could be located anywhere in the process. To maximize throughput, which often has a large economic benefit, the TPM should be located at the production bottleneck so that the bottleneck constraint can be tightly controlled (with minimum "back-off"). 

\begin{figure}[tbh]
\centering
\includegraphics[width=1.0\linewidth]{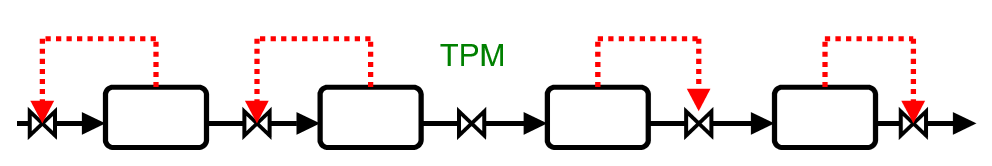}
\caption{Radiation rule for inventory control}
\label{fig:FIG-TPM}
\end{figure}

The inventory control system needs to be consistent, which means that the mass balances are satisfied at all locations at steady state.
We have the following consistency rules regarding inventory control and TPM location:
\begin{itemize}
{\em 
\item Rule C1. One cannot control (set the flowrate) the same flow twice.
\item Rule C2. Controlling the inlet or outlet pressure indirectly sets the in- or outflow (and indirectly makes the MV (valve) used for pressure control a TPM).
\item Rule C3. Follow the {\bf radiation rule} whenever possible.
\item Rule C4. No inventory loop can cross the TPM. }
\end{itemize}
The radiation rule says that, to get a consistent inventory control system using local control loops, the inventory control loops must be radiating around the TPM \citep{Price1993}.
This means that inventory control must be in the direction of flow downstream of the TPM, and opposite the direction of flow upstream the TPM, see Figure~\ref{fig:FIG-TPM}. 

Traditionally, the TPM location and thus the inventory control system is fixed. 
However, a fixed inventory control system means that changes in operation, including reaching new production bottlenecks, temporary stops for maintenance, batch operation, and some changes in feed and product rates, will require manual intervention. This puts a heavy burden on the operators and leads to non-optimal operation.  \cite{Shinskey1981} has proposed a very clever bidirectional control architecture for automatic reconfiguration of inventory control loops which also automatically moves the TPM  to the current bottleneck. 
% and maximize production in response to changes in active constraints.
This approach is next applied to the important industrial case of gas-liquid separation of a two-phase hydrocarbon stream from a well.

\begin{figure}[tbh]
\centering
\includegraphics[width=0.9\linewidth]{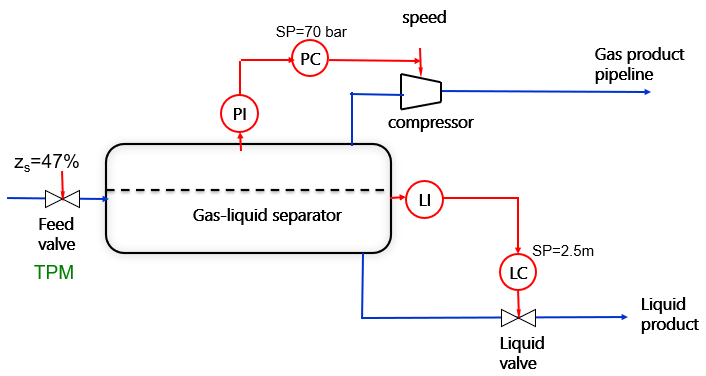}
\caption{Nominal control system for case study I with TPM at feed (no constraints encountered).}
\label{fig:FIG1}
\end{figure}
 
\subsection{Application of bidirectional control to case study I} 
The nominal control system where the operator manually sets the feed rate (choke valve position $z$) is shown in Figure~\ref{fig:FIG1}. In this case, the TPM is at the feed, so inventory control of the downstream separator tank is in the direction of flow: 
The gas inventory is controlled using the pressure controller (PC) with the compressor speed as the manipulated variable (MV), and the liquid inventory is controlled using the level controller (LC) with the liquid outflow valve as the MV. 

\begin{figure}[tbh]
\centering
\includegraphics[width=1.0\linewidth]{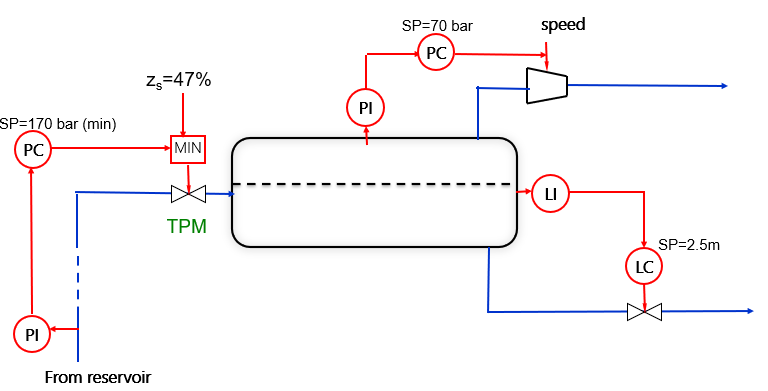}
\caption{TPM at feed with override selector to satisfy minimum well pressure of 170 bar at  reservoir. }
\label{fig:FIG2}
\end{figure}

However, if the operator opens the feed valve $z$ too much, it may happen that
the upstream well pressure becomes too low ($p_{min}=170$ bar) which may damage the reservoir.
This may be solved using a minimum pressure control override as shown in Figure~\ref{fig:FIG2}. 
From Rule C2, since we are setting the inlet pressure, the TPM remains at the feed valve, and the downstream inventory control system for level and pressure is unchanged. 
Note that the MIN-selector allows the operator to reduce the feed rate further when desired by setting a smaller value for $z_s$. 
On the other hand, if the objective is to maximize production, the operator may set $z_s$= 100\% (or, equivalently, the operator may set $z_s=\infty$, since the
valve anyway has a build-in maximum constraint at $z_{max}=100$\%).

If the gas flow from the reservoir becomes too large, then the compressor speed may go it its maximum value (100\%) and the separator pressure can no longer 
be kept at its nominal setpoint of 70 bar.  The solution is to use an "override" where we reduce the feed, that is, 
we let the feed valve take over as the MV for the pressure control.
This is a case of MV-MV switching which may be achieved with either split-range control (one controller and a figure or table with MV switching values) 
or with split-parallel control (two controllers with different setpoints) \citep{Skogestad2023}.
The first option becomes a bit complicated because the switching value for the feed valve is not 100\%, 
but given through the MIN-selector as either the operator setpoint ($z_s$) or the ``override'' value from the reservoir pressure controller (with SP=170 bar).

\begin{figure}[tbh]
\centering
\includegraphics[width=1.0\linewidth]{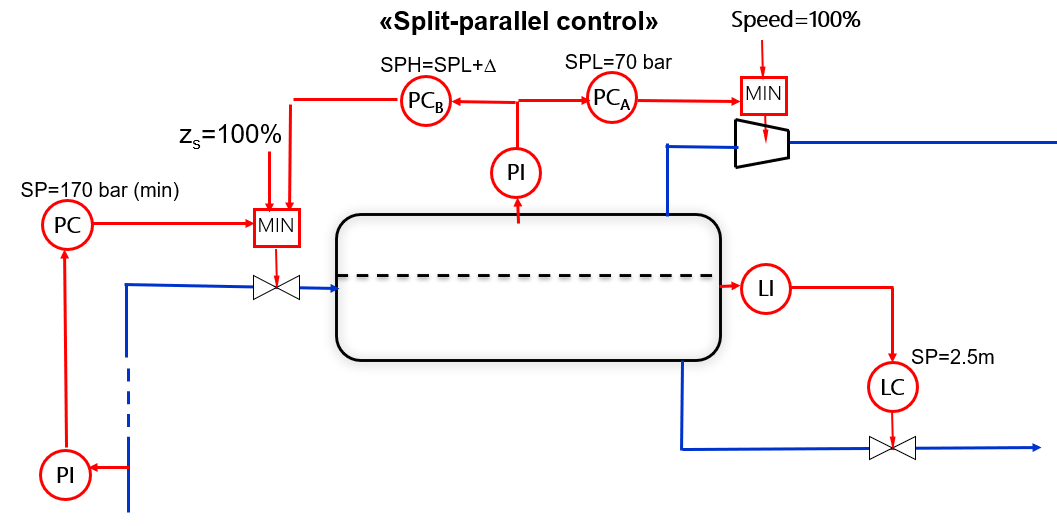}
\caption{Bidirectional control scheme where TPM moves automatically to the bottleneck (feed valve or compressor) to maximize throughput.}
\label{fig:FIG3}
\end{figure}

We therefore use split-parallel control as shown in Figure~\ref{fig:FIG3}, 
which also has the advantage of allowing different controller tunings for the two pressure controllers (PC$_A$ and PC$_B$).
Since the separator pressure will start rising when the compressor saturates at 100\%, the setpoint SPH for PC$_B$ should be higher than SPL for PC$_A$, 
that is, SPH = SPL + $\Delta$. With $\Delta=1$ bar, this gives SPH = 71 bar.

When split-parallel control is combined with MIN-selectors as shown in Figure~\ref{fig:FIG3},
this results in "bidirectional inventory control" \citep{Shinskey1981, Zotica22-bidirectional} of the separator pressure, where the throughput is maximized and the TPM is automatically moved to the present bottleneck (feed valve or compressor).

Note the following about the bidirectional scheme in  Figure~\ref{fig:FIG3}:
\begin{itemize}
 \item We used the term "override" for controller PC$_B$, but this is a bit misleading because this is the optimal thing to do. This is illustrated by the fact that, if we start from a situation where the TPM is at the compressor (e.g., with the compressor speed set at 100\%), then controller PC$_A$ becomes the "override" for pressure control if the feed rate becomes the bottleneck.
\item For the split-parallel control, 
there should normally be a period during the MV-MV switching (between the SPL and SPH setpoints) where none of the two controllers (PC$_A$ and PC$_B$) are active, so the controlled variable (pressure in this case) is ``drifting''.
%because both MVs are either saturated or set by another controller.
%(for example, the compressor speed is at 100\%  and the feed valve $z$ is set at a low value by the reservoir pressure controller). 
However, if one uses a too small setpoint difference $\Delta$, then 
%there may be a period during the MV-MV switching when 
both PC$_A$ and PC$_B$ (which control the same pressure) may be active at the same time during the MV-MV switching and start fighting. 
This may slow down the time for switching back when we go out of a constraint (e.g., the compressor is no longer a bottleneck) and we want to increase the throughput. This is related to an internally unstable mpde \citep{Forsman2025}. Thus, it is possible that a setpoint difference $\Delta= 1$ bar is too small. 
A larger setpoint difference also has the advantage of delaying the switch between using the compressor and feed valve as MVs for separator pressure control, which may be an advantage if the constraints are only temporary. 
\end{itemize}

\begin{figure}[tbh]
\centering
\includegraphics[width=1.0\linewidth]{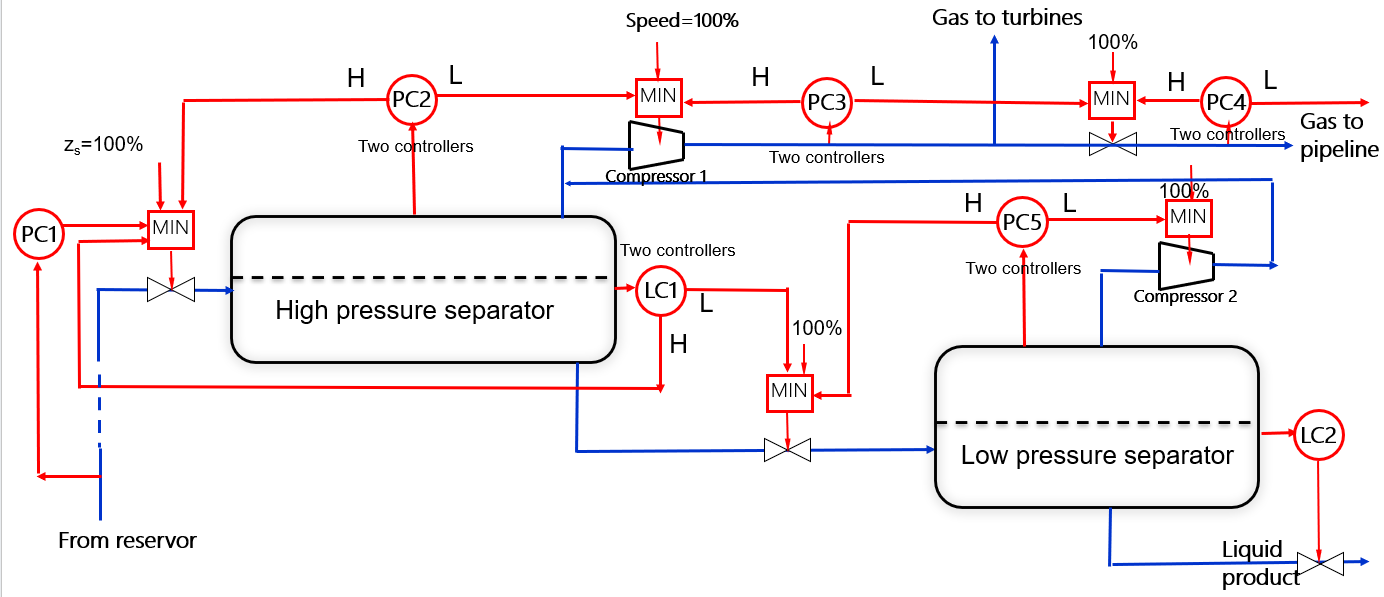}
\caption{Final bidirectional control scheme for case study I for more complex process with additional constraints. The TPM can be at all 5 MIN-selectors. }
\label{fig:FIG4}
\end{figure}

The bidirectional inventory control idea may be extended to the more complex case in 
Figure~\ref{fig:FIG4}, where we have added a low-pressure separator with
another compressor, a gas turbine and additional constraints. In Figure~\ref{fig:FIG4}, we also use a simplified notation where the two split-parallel controllers (e.g.., PC$_A$ and PC$_B$ in Figure~\ref{fig:FIG3})
are combined into one controller block (e.g., PC2) with two setpoints L=SPL and H=SPH, where H=L+$\Delta$.

%FIG.4: Have added another low-pressure tank (which releases more gas) and more constraints. 

In summary, the following seven constraints are handled in Figure~\ref{fig:FIG4}:

\begin{enumerate}
\item Min. reservoir pressure: "Override" from PC1 using a MIN-selector.
\item Saturation (max. speed) for Compressor 1: PC2-H reduces the feed rate (after some transition time from setpoint L to H). 
\item Saturation (max. valve opening) for liquid leaving separator: LC1-H  reduces the feed rate (also here after some transition time from setpoint L to H).
\item Saturation (max. speed) for Compressor 2: PC5-H reduces the liquid feed. This results in loss of level control (LC1-L), so LC1-H will reduce the feed rate.
\item Min. pressure for gas to turbines: PC3-L closes the extra gas valve (after the split).
\item Max. pressure at inlet gas pipeline: PC4-H closes the same extra gas valve.
\item Min. pressure at inlet gas pipeline. PC4-L closes a downstream valve (not shown in Figure~\ref{fig:FIG4}).
\end{enumerate}

Note the following in Figure~\ref{fig:FIG4}:

\begin{itemize}
\item The bidirectional scheme automatically moves the TPM (after some transition time) to the MV-location (valve/compressor) determined by which of the above seven constraints is active.  
\item All MIN-selectors have an external setpoint, which allows the operator to reduce the flow and move the TPM to this location, if desired. In Figure~\ref{fig:FIG4}, the setpoints are set at 100\% because the objective is to maximize throughput. More generally, the external setpoint could be the optimal unconstrained value (without constraints).
\item All controllers reduce the MV value (valve/speed) when activated (and do the opposite when deactivated).
\item Note that all the constraints are consistent in that they all are satisfied (in the end) by reducing the feed rate. Thus, we only need MIN-selectors.
\item We have H=L+$\Delta$ for the setpoints of the five split-parallel controllers. Note that with a larger setpoint separation $\Delta$, the buffer of gas or liquid between  setpoints L and H (while no control loops are active) will delay the transition. Thus, it will take more time before the feedrate needs to be reduced, which is an advantage economically. 
\item The inlet pressure to the exit gas pipeline is often "floating", that is, uncontrolled, which means that the valve between PC3 and PC4 is fully open (as in Figures~\ref{fig:FIG1}-~\ref{fig:FIG3}) which is an economic advantage as it reduces compression power. In this case, the values for H and L for PC4 may be set far apart, for example, H=200 bar and L=100 bar.
\end{itemize}

The solution in Figure~\ref{fig:FIG4} may at first seem complex, but note that the structure is repeating: all valves/compressors have MIN-selectors with an L-setpoint from the upstream unit, an H-setpoint from the downstream unit and a desired setpoint (which can be given up; in  Figure~\ref{fig:FIG4} it is 100\% because we want to maximize production). Correspondingly, all inventory controllers (LC and PC) have an H-setpoint to the upstream valve/compressor and an L-setpoint to the downstream valve/compressor. 

\section{Case study IIA: Happy cows}

\begin{figure}[tbh]
\centering
\includegraphics[width=1.0\linewidth]{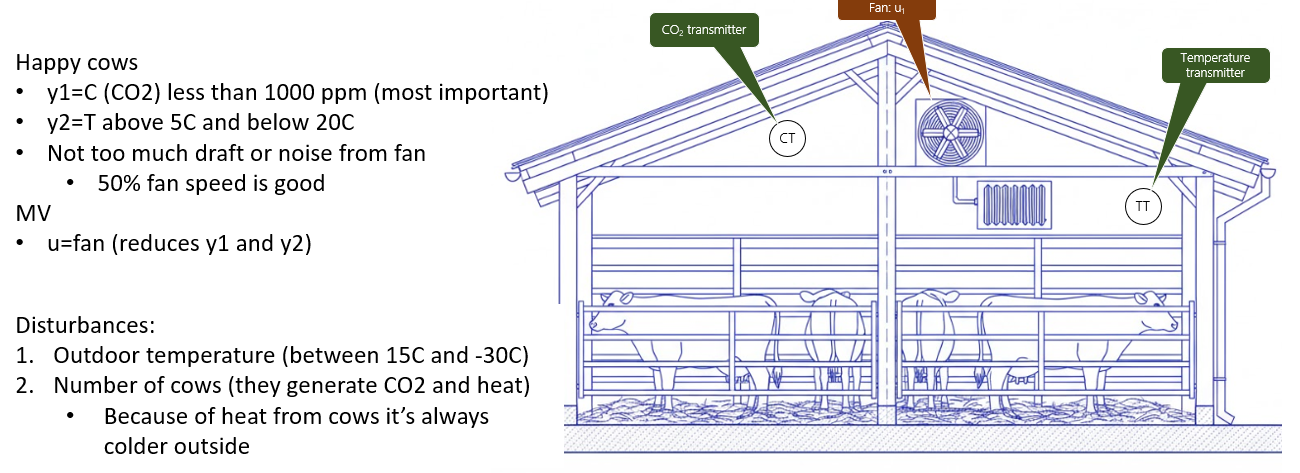}
\caption{Case study IIA: Barn with fan speed as MV}
\label{fig:COW1}
\end{figure}

Consider the barn for cows in Figure~\ref{fig:COW1} with the fan speed as the manipulated variable (MV). To keep the cows happy, the air quality should be good with CO2-concentration ($y_1=c$) below 1000 ppm, and the temperature $y_2=T$ in the barn should be between 5C and 20C. 
Nominally (and hopefully most of the time), the fan speed $u$ will be fixed at the desired value (nominal setpoint) of 50\%. However, if there are too many cows, the air quality may get bad (with CO2 rising above 1000 ppm) or the temperature may get too high (above 20C) and we need to increase the fan speed above 50\%. On the other hand, on a cold winter day, the temperature may get too low (below 5C) and we need to reduce the fan speed below 50\%.  
%Note that the air quality constraint ($y_1<1000$ ppm) is more important than the temperature constraint ($y_2>5$C).

In summary, there are three control objectives (constraints) (one on $y_1$ and two on $y_2$) and only one manipulated variable ($u$=fan speed). We will use one controller for each of the three constraints and selectors on the three controller outputs so that only one constraint is actively controlled at a time. 
\begin{itemize}
{\em
\item Selector Rule S1: In general, we need a MAX-selector for constraints that are satisfied by a large value of the input $u$, and a MIN-selector for constraints that are satisfied by a small value of the input $u$ \citep{Skogestad2023}. }
\end{itemize}
From selector rule S1, we have for the three constraints:

\begin{enumerate}
\item Constraint $y_1=c \le 1000$ ppm: Satisfied by large fan speed $\Rightarrow$ MAX-selector
\item Constraint $y_2=T \ge 5$C: Satisfied by small fan speed $\Rightarrow$ MIN selector
\item Constraint $y_2=T \le 20$C: Satisfied by large fan speed $\Rightarrow$ MAX selector
\end{enumerate}

\begin{figure}[tbh]
\centering
\includegraphics[width=0.8\linewidth]{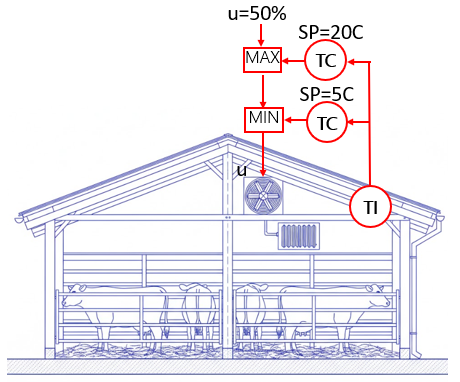}
\caption{Case study IIA: Control of temperature within a band using the fan. Since the constraints ($T\geq 5$C, $T\leq 20$C) never conflict, the order of MAX- and MIN-selectors may be interchanged, or replaced by a MID-selector.}
\label{fig:COW2A}
\end{figure}

\begin{figure}[tbh]
\centering
\includegraphics[width=0.8\linewidth]{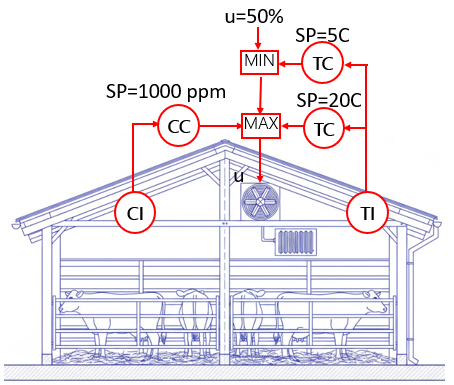}
\caption{Case study IIA: Final proposed structure with control of both CO2 and temperature. The MAX-selector is at the end because max. CO2 (at 1000 ppm) has higher priority than min. $T$ (at 5C).}
\label{fig:COW2}
\end{figure}

In general, when we require both MAX- and MIN-selectors for an input (here, for $u$=fan speed), there may be cases where the constraints are conflicting
and there is no feasible solution. 
\begin{itemize}
\item {\em Selector Rule S2: In cases where both a MAX- and MIN-selector is needed, and there is a potential conflict, we must decide which constraint is the most important and put the corresponding selector “at the end” of the sequence (closest to the MV (input) $u$).}
\end{itemize}
In our case, the MIN- and  MAX-selectors for constraints 2 and 3 will never conflict because they are on the same variable, $y_2=T$, and temperature cannot be both 5C and 20 C at the same time. A feasible control structure for this band temperature control is shown in Figure~\ref{fig:COW2A}. Here, the MIN-selector is last (at the end), but the order could be interchanged. This is shown in Figure~\ref{fig:COW2} where the order matters because constraint 1 on CO2 is included.
%However, the order of the selectors matters when we include the CO2-constraint. When it is cold outside and many cows, constraint 1 on CO2-level (max. 1000 ppm) (requiring a large fan speed and a MAX-selector) may conflict with constraint 2 on temperature (min. 5C) (requiring a small fan speed and a MIN-selector). Since constraint 1 on CO2 %(max. 1000 ppm CO2) 
%(requiring a MAX-selector) has higher priority (see Figure~\ref{fig:COW1}),
%than the minimum temperature constraint of 5C (requiring a MIN-selector) , 
%the MAX-selector should be at the end. 
Since the CO2-constraint has higher priority, we may need to accept that $y_2=T$ drops below 5C when it is very cold outside.

The proposed control structure in Figure \ref{fig:COW2} has three controllers and two selectors. 
It is suggested to use PI-controllers, and because at most one controller is active at a time, all three controllers need to have anti windup on the integral action. 
With the given constraints and priorities, the proposed control structure will be optimal at steady state, and also the dynamic responses are good (as confirmed by simulations). 
MPC is an alternative, but it requires a dynamic model and online optimization and it seems unnecessary complicated.  

Note that we have put a nominal desired value $u=50$\% into the first selector block in Figures~\ref{fig:COW2A} and \ref{fig:COW2}. More generally, we have:
\begin{itemize}
\item
{\em Selector Rule S3: It is recommended to always put  a "desired input" $u=u_0$ (50\% in Figures~\ref{fig:COW2A} and \ref{fig:COW2}) into the first selector, and sometimes into selectors later in the sequence. 
Otherwise, there may not be a unique solution in cases when no constraints are active.
Here, $u_0$ could be a given desired value (e.g., set by the operator), an optimized value (resulting from an unconstrained optimization), 
or the output from a controller which has a setpoint which can be given up.

Comment: In some cases, the desired $u_0$ is the maximum value for the MV (100\%) (e.g., maximize production), which is relevant for a MIN-selector, and in this case  we may omit the desired value $u_0=100$\% (and still get a unique solution) because all MVs (valves) have a built-in MIN-selector at $u_{max}=100$\% (selector Rule 3 in \cite{Skogestad2023}).  
Nevertheless, it may be good to keep the desired value $u_0=100$\% (or $u_0=\infty$) for clarity. An example is given in  Figure~\ref{fig:FIG3} on bidirectional control for case study I. 

In other cases, the desired $u_0$ is to use the minimum input (0\%), which is relevant for a MAX-selector, and in this case  we may omit the specification $u_0=0$\% (and still get a unique solution) because all MVs (valves) have a built-in MAX-selector at $u_{min}=0$\%.  
Nevertheless, it may be good to keep the desired input $u_0=0$\% for clarity. }

\end{itemize}

\section{Case study IIB:  Avoiding unhappy cows}
In this extension of case study IIA, we are more worried about cold days, so we add a heater (which is rather expensive to use) as an MV. 
However, on really cold days, the heater may saturate at maximum, and to avoid unhappy freezing cows (with $y_2=T$ below 0C), we will allow reducing the fan speed and let the CO2 level go above 1000 ppm. However, to avoid very unhappy (sick) cows, we cannot let the CO2 level go above 3000 ppm. The specifications for case study IIA are summarized in Table~\ref{tab:cows}.

%In summary, we have for the even more happy cows in case study IIB

\begin{table}[htb]
\centering
\caption{Cows' comfort conditions and manipulated variables for Case study IIB.}
\label{tab:cows}
\begin{tabular}{ll}
\hline
\multicolumn{2}{l}{\textbf{Happy cows}} \\
\hline
$y_1 = c_{CO2}$ & less than 1000 ppm \\
$y_2 = T$ & between 5\,$^\circ$C and 20\,$^\circ$C \\
Nominal (desired) & 50\% fan speed \\[4pt]
\multicolumn{2}{l}{\textbf{Unhappy (freezing) cows}} \\
\hline
$y_2 = T$ & less than 0\,$^\circ$C \\[4pt]
\multicolumn{2}{l}{\textbf{Very unhappy (sick) cows}} \\
\hline
$y_1 = c_{Co2}$ & above 3000 ppm \\[4pt]
\multicolumn{2}{l}{\textbf{MVs (inputs)}} \\
\hline
$u_1 =$ fan speed & decreases $y_1$ and $y_2$ (cheap) \\
$u_2 =$ heater & increases $y_2$ (expensive) \\
\hline
\end{tabular}
\end{table}

\begin{figure}[b]
\centering
\includegraphics[width=0.8\linewidth]{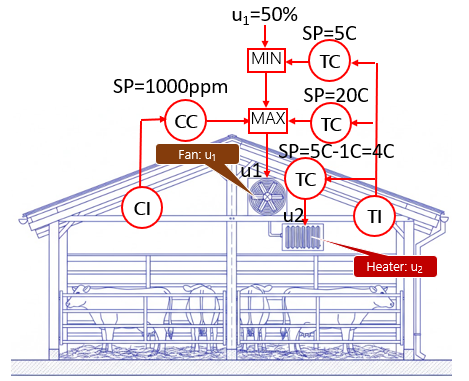}
\caption{Intermediate control structure for case study IIB with addition (compared to Figure~\ref{fig:COW2}) of split-parallel control using heater u2.}
\label{fig:COW3A}
\end{figure}

\begin{figure}[bth]
\centering
\includegraphics[width=0.8\linewidth]{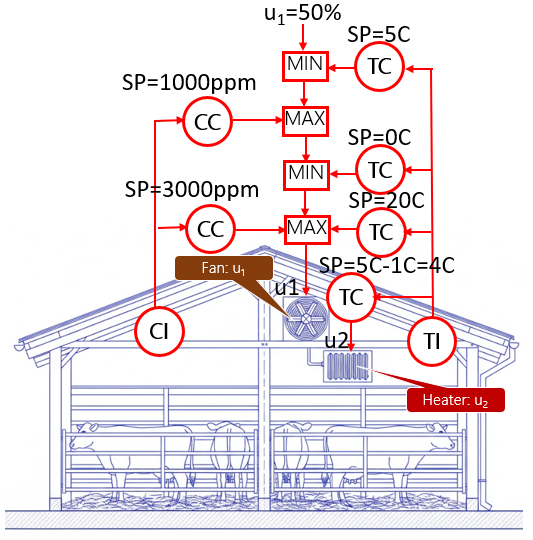}
\caption{Final control structure for case study IIB: With 
%both fan and heater as MVs and 
additional constraints on CO2-level (3000 ppm) and temperature (0 C) to avoid unhappy cows.}
\label{fig:COW3}
\end{figure}

The final proposed  control structure is shown in Figure~\ref{fig:COW3} and its design is explained in the following.

{\bf Split-parallel control using heater}. First, note that both manipulated variables (MVs) $u_1$ and $u_2$ affect the temperature, but we only want to manipulate one MV at a time for temperature control.
This is a case of MV-MV switching which may be achieved with split-range control (one controller plus a figure or table with MV switching values) 
or split-parallel control (two controllers with different setpoints) \citep{Skogestad2023}.
The first option becomes a bit complicated because the switching value for the fan is not 0\%, 
but given by a varying ``override'' value from the CO2-controller (SP=1000 ppm) through the MAX-selector.
We therefore use split-parallel control, which also has the advantage of allowing different controller tunings for the two TCs.
Since reducing the fan speed ($u_1$) is the cheap alternative to keep the temperature above 5C, it should be used first, with a setpoint of 5C. 
The setpoint for the heater ($u_2$) is then 5C-$\Delta $, where we choose a setpoint difference $\Delta=$1C. 
The heater will then activate on cold days when the CO2-controller with SP=1000 ppm is using the fan (through the MAX-selector), 
and the temperature drops to 4C.
The proposed intermediate control structure with additional of split-parallel control is shown in Figure~\ref{fig:COW3A}.

% Does not seem to likely, depends on tracking for TC with SP=5C:
%One has to be careful about using a too small setpoint difference $\Delta$, because otherwise the two controllers (with setpoints 5C and 4C) may start fighting and slow down the transition time for switching back from expensive heating to cheaper fan when the outdoor temperature increases again (and maybe some cows have gone outside so the fan speed is 50\%).

{\bf Really cold days with additional constraints.} The heater is not very large so if it gets really cold outside it may saturate at maximum (100\%) and temperature control is lost and the temperature drops below 4C. However, if it reaches 0C, we override the CO2-controller (and give up keeping SP=1000 ppm) by reducing the fan speed with a MIN-selector. 
The CO2-concentration is then uncontrolled, but we cannot allow the CO2-level to become too high because the cows get sick. If it reaches 3000 ppm, we override the temperature control (and give up keeping SP=0C) by increasing the fan speed with a MAX-selector. 
In this case, temperature will drop below 0C.
%, so we probably should cover the cows with blankets (as a third manual MV). Similarly, on hot days where the temperature exceeds 25C we may spray water on the cows (as a fourth manual MV).
 
The final control structure in Figure~\ref{fig:COW3} may seem complicated, but note that we are using one controller for each constraint, with the 
highest priority (and non-conflicting) constraints located at the end of the sequence, closest to the input $u_1$ (fan speed).

\section{Discussion}

%The advanced regulatory control (ARC) PID schemes presented in the case studies in this paper may at first seem complicated, especially for engineers with little training on such solutions. Actually, from industrial experience, it seems that operators may accept these control structures more easily than engineers because the way they work resemble the manual overrides the operators need to do when new constraints are encountered. 

Simulations show that the proposed PID-based control strategies work well. Nevertheless, there are many theoretical issues to be resolved about the dynamic behavior and stability of switching schemes.
%, such as the one shown for case study IIB in Figure~\ref{fig:COW3}. 
For example, for certain parameter values and noisy measurements, one may get chattering (e.g., \cite{Caponigro2018}) and even chaotic behavior. This complicates the theoretical analysis, and may be one reason why there has been little work in academia on such schemes. In practice, industrial experience over more than 60 years \citep{Maarleveld1970,Smith10}, have shown that with proper design, the selector-based swithcing schemes work well with smooth transitions between changing active constraints. In some cases, to avoid chattering, one may put restrictions on the switching frequency (e.g., dead-band or dwell time) but with a good anti-windup scheme (e.g., the tracking scheme of Astrom described in \cite{Skogestad2023}) it is usually not necessary.  The simulations use this scheme, with the tracking time $\tau_T$ set equal to the integral time $\tau_I$, and provide smooth transitions between changes in active constraints.

Although a hierarchical (and decentralized) switching network of PID controllers may be used in complex cases (case study IIB), there are certainly other cases where it does not apply, and where model-based schemes such as MPC/RTO will be a better, and sometimes even simpler, solution.  Specifically. the switching schemes presented in this paper, assume that one makes a pairing between each constraint and a single manipulated variable (input), and with multiple constraints (associated with one input) requiring both a MAX- and MIN-selector, this may result in an apparent infeasibility, which could have been avoided with a multivariable scheme.   
\cite{Bernardino2024} show that if we make the restrictive assumption that we have at least as many inputs as constraints (which is restrictive and not satisfied for case study IIB) then it is possible to find a decentralized switching scheme that achieves optimal operation at steady state. Rules and recommendations for more general cases with many constraints are lacking.

\section{Conclusion}

In the academic community, there is a myth that you need a detailed nonlinear model and an on-line optimizer (RTO) if you want to optimize the process, and you need a dynamic model and model predictive control (MPC) if you want to handle constraints. The objective of this paper, is to show by some simple case studies that this is not true:
Because the most important for optimal operation is usually to implement the active constraints, one can in many cases obtain optimal operation by measuring and controlling the constraints.  This is fairly obvious, but what is probably less clear is that the MIN-selectors in the ARC schemes also switch back, that is, they stop controlling the constraint when the disturbance goes away.  
%ARC is realized using PID control together with some logic (e.g., selectors and split-parallel control). 
%Because most industrial ARC-applications seem {\em ad hoc} and few systematic design methods exist, this fact is not well known, even to control professors. 
Today advanced PID solution (ARC) and model-based control and optimization (MPC/RTO) are in parallel universes, but both solutions are needed in the future control engineer's toolbox.

{\em Simulations and Matlab/Simulink files for the case studies are available on the home page of the author.}\\
\texttt{\scriptsize https://skoge.folk.ntnu.no/publications/2026/skogestad-PID-ifac2026/}

\begin{ack}
Discussions with Krister Forsman and Mohammed Adlouni are gratefully acknowledged.
\end{ack}

\bibliography{ifacconf}

\section*{Appendix: Simulations}

\subsection*{\bf Case study IIB: Avoiding unhappy cows}

We model the barn as a single well-mixed continuous-stirred-tank reactor (CSTR)
with two states $(c, T)$. Here $c$ [m$^3$ CO2/m$^3$] is the volumetric (or mole) fraction of CO2, usually given in ppm (so multiplied by 10$^6$). 
The mass balance for CO\textsubscript{2} and the energy
balance are
\begin{align}
V\,{dc\over dt} \;=\;& N_{\text{cows}}\,G_{\text{CO}_2}  + q_{\text{air}}(u_1)(c_{\text{out}} - c),
\label{eq:co2balance}\\[2pt]
\rho c_p V\,{dT\over dt} \;=\;&
N_{\text{cows}} Q_{\text{cow}} + Q_{\text{h,max}}\frac{u_2}{100\%} \nonumber \\ 
& - \left(\rho c_p q_{\text{air}}(u_1) + UA \right) (T - T_{\text{out}})
\label{eq:Tbalance}
\end{align}
where we have for the fan
\begin{equation}
 q_{\text{air}}(u_1) \;=\; q_{\min} + (q_{\max} - q_{\min})\,\frac{u_1}{100\%}.
\label{eq:fanmap}
\end{equation}
Numerical values are listed in
Table~\ref{tab:plant-params}. 
%; they were chosen as plausible defaults for a medium Norwegian dairy barn, since the original paper does not provide values.

\begin{table}[!ht]
\centering
%\hline
\caption{Parameters for cow case study II.}
\label{tab:plant-params}
\begin{tabular}{l l l}
\hline % \toprule
\textbf{Symbol} & \textbf{Value} & \textbf{Description} \\
\hline % \midrule
$V$                          & 3\,000\,m\textsuperscript{3}              & Air volume of the barn \\
$N_{\text{cows}}$ (nominal)  & 80                                       & Cow occupancy \\
$G_{\text{CO}_2}$            & $5\times10^{-5}\,{m^3/s\over cow}$ & Per-cow CO\textsubscript{2} generation \\
$Q_{\text{cow}}$             & 1\,000\,W/cow                            & Per-cow metabolic heat \\
$q_{\max}, q_{\min}$         & 15, 0.1\,m\textsuperscript{3}/s          & Fan airflow at 100\,\% and 0\,\% \\
$Q_{\text{h,max}}$               & 50\, 000 \, W                                  & Heater rating at 100\,\% \\
$UA$                         & 2\, 000 W/K                              & Barn heat-loss coefficient \\
$c_{\text{out}}$             & 420\,ppm                                 & Outdoor CO\textsubscript{2} concentration \\
$T_{\text{out}}$ (nominal)   & 0\,\textdegree C                         & Outdoor temperature \\
$\rho$                & 1.2\,kg/m\textsuperscript{3} & Air density \\
$c_p$                & 1\,005\,J/kg/K & Air specific heat capacity\\
\hline % \bottomrule
\end{tabular}
\end{table}

{\bf Simulations}

\begin{figure}[bth]
\centering
\includegraphics[width=\columnwidth,trim=250 0 0 70,clip]{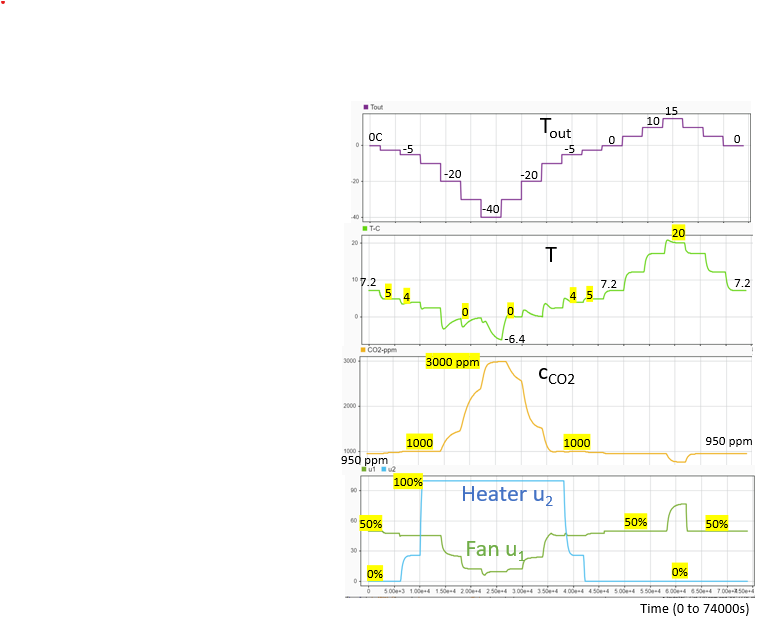}
\caption{Dynamic simulation of final control structure in Figure~\ref{fig:COW3} for cow case study IIB. Step changes in $T_{out}$ occur every 4000 s.}
\label{fig:COWSIM}
\end{figure}

We simulate for a range of step changes in outdoor temperatures ($T_{out}$) occuring with intervals of 4000s, giving a total simulation time of 74000 s = 20.5h. We start from from $T_{out}=0$C, and step gradually down to -40C (which may seem extreme, but it may happen in Norway), then gradually up to 15C, and then back again to the nominal point with $T_{out}=0$C.The dynamics responses are good with smooth switching as shown in Figure~\ref{fig:COWSIM}. The responses are almost the same if we add time delays of 1 min for each measurement ($T$ and $c_{CO2}$), and also time delays of 3 min yield only minor oscillations. Note that all controllers are active in some period, and we reach (or at least approach) the steady states given in Table~\ref{tab:operating_points}.  Note that there are two degrees of freedom ($u_1$ and $u_2$) so at each steady state there are two active constraints or active setpoints
\setlength\fboxsep{1pt}  % for yellow box
(in \colorbox{yellow!35}{yellow}). 
Note that outdoor temperatures above about 18C (not simulated) result in \colorbox{yellow!35}{$u_1=100\%$} (max. fan) and \colorbox{yellow!35}{$u_2=0\%$} (no heat, of course), and the indoor temperature $T$ exceeds the bound of 20C. 

Comment: For this reason, on hot days, where the indoor temperature exceeds 20C, we may spray water on the cows (as a third MV) to keep the cows happy. Similarly, we may cover the cows with blankets (as a fourth MV) if the indoor temperature drops below 0C. 

\setlength\fboxsep{1pt}  % for yellow box
\begin{table}[t]
\centering
\setlength{\tabcolsep}{4pt}
\caption{Steady-state operating points for cow case study IIB 
\\
\small{The active constraints/setpoint are in \colorbox{yellow!35}{yellow}.}
}
\label{tab:operating_points}
\begin{tabular}{c|cccc}
\hline
$T_{\mathrm{out}}$ & $T$ & $c_{CO2}$ & Fan $u_1$ & Heat $u_2$ \\
($^\circ$C) & ($^\circ$C) & (ppm) & (\%) & (\%) \\
\hline
15   & \colorbox{yellow!35}{20.0} & 765  & 77.2 & \colorbox{yellow!35}{0}   \\
10   & 17.2 & 950  & \colorbox{yellow!35}{50.0} & \colorbox{yellow!35}{0}   \\
5    & 12.2 & 950  & \colorbox{yellow!35}{50.0} & \colorbox{yellow!35}{0}   \\
0 \tiny{(nominal)}   & 7.2  & 950  & \colorbox{yellow!35}{50.0} & \colorbox{yellow!35}{0}   \\
-2.5 & \colorbox{yellow!35}{5.0} & 977  & 47.6 & \colorbox{yellow!35}{0}   \\
-5   & \colorbox{yellow!35}{4.0} & \colorbox{yellow!35}{1000} & 45.6 & 25.7 \\
-10  & 2.6  & \colorbox{yellow!35}{1000} & 45.6 & \colorbox{yellow!35}{100} \\
-20  & \colorbox{yellow!35}{0.0} & 1492 & 24.4 & \colorbox{yellow!35}{100} \\
-30  & \colorbox{yellow!35}{0.0} & 2487 & 12.3 & \colorbox{yellow!35}{100} \\
-40  & -6.4 & \colorbox{yellow!35}{3000} & 9.7  & \colorbox{yellow!35}{100} \\
\hline
\end{tabular}
\end{table}

\begin{table}[h]
\centering
\caption{PI tunings used in simulation of cow case study IIB.}
\begin{tabular}{l l l l c}
\hline
 & MV & $K_c$ & $\tau_I$ & Setpoint \\
\hline
TC1 & Fan ($u_1$) & $-10\ \%/^\circ\mathrm{C}$ & $350\ \mathrm{s}$ & 20C \\
TC3 & Fan ($u_1$) & $-10\ \%/^\circ\mathrm{C}$ & $350\ \mathrm{s}$ & 5C \\
TC2 & Fan ($u_1$) & $-3.33\ \%/^\circ\mathrm{C}$ & $1050\ \mathrm{s}$ & 0C \\
CC2 & Fan ($u_1$) & $-0.1\ \%/\mathrm{ppm}$ & $350\ \mathrm{s}$ & 1000 ppm \\
CC1 & Fan ($u_1$) & $-0.02\ \%/\mathrm{ppm}$ & $1750\ \mathrm{s}$ & 3000 ppm \\
TC  & Heater ($u_2$) & $22\ \%/^\circ\mathrm{C}$ & $350\ \mathrm{s}$ & 4C \\
\hline
\end{tabular}
\label{tab:PItunings}
\end{table}

{\bf Linearized model analysis}

Note that the two balances \eqref{eq:co2balance} and \eqref{eq:Tbalance} are decoupled, that is, without control, $c$ (CO2) is not affected by changes in $T$, and $T$ is not affected by changes in $c$.  
The two time constants at the nominal operating point (where $u_1=50\,\%$ giving   $q=7.55$ m$^3$/s) are
\begin{equation}
\tau_c = \frac{V}{q} \approx 387\text{s}, \ 
\tau_T = \frac{\rho c_p V}{\rho c_p q + UA} \approx 334\text{s}
\label{eq:taus}
\end{equation}
The steady-state linearized gains from $u_1$ (fan speed) to $c$ and $T$ are at the nominal operating point (where $u_1=50\,\%$ and $u_2=0 \%$ giving  $q=7.55$ m$^3$/s and $T=7.2$ C)
\begin{equation}
k_{c,u_1} = 
-\frac{N_{\mathrm{cows}} G_{\mathrm{CO2}}
(q_{\max}-q_{\min})}
{q^2} \, \frac{10^6}{100\% } \approx -10.5 \frac{\text{ppm}}{\%} 
\label{eq:cgain1}
\end{equation}
\begin{equation}
k_{T,u_1} = 
-\frac{(T-T_{\mathrm{out}}) \rho c_p}{\rho c_p q + UA} \frac{(q_{\max}-q_{\min})}{100\% } \approx -0.12 \frac{\text{C}}{\%} 
\label{eq:Tgain1}
\end{equation}
Similarly, from $u_2$ (heater) to $T$:
\begin{equation}
k_{T,u_2} = \frac{Q_{h,\max}/100\%}
{\rho c_p q + UA} \approx 0.045  \frac{\text{C}}{\%} 
\label{eq:Tgain2}
\end{equation}

Note that the fan flowrate $q$ may vary a lot. It is nominally 7.55 m3/s but it may drop to  2 m3/s or lower when it is extremely cold (-40 C) outside. This means that both the time constants and gains may vary a lot. Fortunately, the initial gains (slopes), $k'=K_c/\tau_c$ and $k'=K_T/\tau_T$  , which are important for controller tuning, vary less.

{\bf PI tunings for cow barn case study}. 

The SIMC PI-tunings if we assume that there is no time delay are
$$ K_c = {\tau\over k \tau_c}  = {1\over k' \tau_c }$$
$$ \tau_I = \min\{\tau, 4 \tau_c\}$$
where the tuning parameter $\tau_c$ is the desired closed-loop time constant.
%For $\tau_I$ we have assumed that $\tau < 4 \tau_c$.
To simplify, we set the nominal time constant for both the CO2 and temperature dynamics equal to $\tau=350$s.
Then, selecting $\tau_c  = 350$s, gives the SIMC tunings in Table~ \ref{tab:PItunings}. 
In addition, all controllers have anti-windup with tracking, using tracking time constant $\tau_T = \tau_I$.

Here are some details about the controller tuning:
\begin{itemize}
\item  Fan ($u_1$) for temperature control:
% k'=-3.6e-4 = -(T-Tout)(qmax-qmin)/V*100.
For TC3 (SP=5C) and TC1 (SP=20C) we use the nominal operating point. This gives $\tau_c=\tau=350$s, $K_c = {1 \over -0.12} = -8.3 \approx -10 $ (rounded up a little for simplicity) and $\tau_I = \tau = 350$s.  
Controller TC2 (SP=0C) is only active when it's very cold outside, and here both the initial slope $k'$ and time constant $\tau$ may be up to a factor 5 larger, which means that $K_c$ should be decreased by a factor 5 and $\tau_I$ increased by a factor 5. However, simulations show that this gave a rather slow approach to the setpoint (0C) at moderately cold days (say, -20 C outdoor), so in the simulations we use a factor 3, which gives $K_c=-3.33$ and $\tau_I=1050$s.  
\item  Fan ($u_1$) for composition (CO2) control: 
%k' = 0.027 = Ncows* GCO2* (qmax-qmin) / (V*q) *e6/100.
For CC2 (SP=1000 ppm) we use the nominal operating point. This gives $K_c = {1 \over -10.5} = -0.095 \approx -0.1 $ (rounded up a little for simplicity) and $\tau_I = \tau = 350$s. 
Controller CC1 (SP=3000 ppm) is only active when it's very cold outside, and here both $k'$ and $\tau$ are about a factor 5 larger, so we choose $K_c=-0.1 /5 = -0.02$ and  $\tau_I = 350\cdot 5 = 1750$s.
\item Heater ($u_2$) for temperature control, TC (SP=4C):  
%k' = 0.00013 = Qhmax/rho*cpV*100 (constant!), 
We get $K_c = {1/0.045} = 22$ and $\tau_I = \tau = 350$s. Note here that  the initial slope $k'$ does not change with the operating point (so the SIMC-value for $K_c$ is independent of $q$ and $T_{out}$).

\end{itemize}
% OLD:
%Fan for T. k=-0.12, tau=334, k'=-36e-4 = -(T-Tout)(qmax-qmin)/V*100. tauc=334 s, 
%tuning: Kc =-8, taui=334  (I used Kc=-10, taui=350s but it oscillates with delay = 60s.) 
%comment: (T-Tout) varies from 7 to 35 (so k' increases by factor 5 on cold days!). So to be %safe reduce Kc by factor 5 when Ts=0. (tau also increases, so I increase taui by factor 2)\\
%Heater for T. k=0.045, tauT=334s. k' = 0.00013 = Qhmax/rho*cpV*100 (constant!), 
%tuning: Kc=1/k=22, taui=334. I use J`Kc=22, taui=350.
%but tauT varies. Anyway, there is no problem. Initial slope is constant and heater is active %(for control) only when u1 is around 40\%. \\
%Fan for CO2. k=-10.5, tauc=387. k' = 0.027 = Ncows* GCO2* (qmax-qmin) / (V*q) *e6/100.
%tunung, Kc=1/k=-0.1 , taui=350. 
%But k' and tauc varies (both increase when q is small, so Kc shopuld decrease and tauI increase, so change both by factor 5 at SP=3000 ppm)
\subsection*{\bf Case study I: High-Pressure Separator}
\label{app:separator-model}

This appendix gives the nonlinear dynamic model and simulations for
 Figure~\ref{fig:FIG3} in Case Study~I. The process is a horizontal high-pressure gas--liquid separator
operating nominally at \(70~\mathrm{bar}\). The separator has two dynamic states:
the liquid level \(h\) and the separator pressure \(P\). 
For control, we consider three controllers: LC, PCA and PCB (the PC for the reservoir pressure is not included, that is, the minimum pressure constraint of 170 bar is assumed to be inactive).  The manipulated inputs for the three controllers are  the liquid-outlet-valve opening \(z_{\mathrm{L}}\), the compressor speed \(z_{\mathrm{G}}\). and the feed-valve opening \(z_{\mathrm{feed}}\) - all in the range 0\% to 100\%.

\begin{figure}[bth]
\centering
\includegraphics[width=\columnwidth]{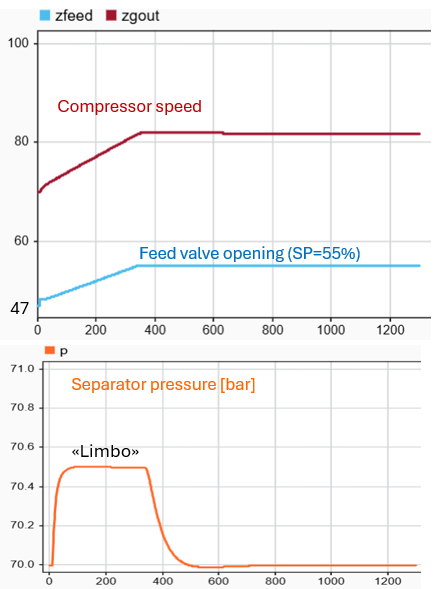}
\caption{Separator with split parallel control (Figure~\ref{fig:FIG3}): Dynamic simulation of setpoint increase in feed valve opening from 47\% to 55\%. Controller PCB keeps back the increase if the feed valve opening, resulting in a ``limbo'' state. (Time is in seconds in all simulations)}
\label{fig:SEPSIM1}
\end{figure}

\begin{figure}[bth]
\centering
\includegraphics[width=\columnwidth]{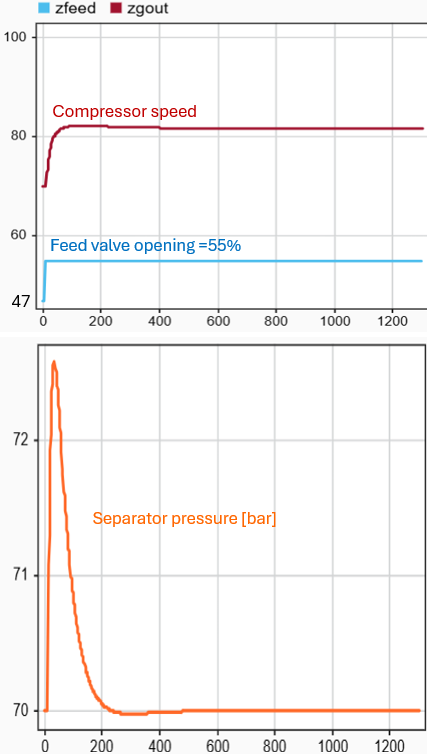}
\caption{Separator (Figure~\ref{fig:FIG3}) with pressure controller PCB deactivated (similar to split range control): Dynamic simulation of increase in feed valve opening from 47\% to 55\% }
\label{fig:SEPSIM1A}
\end{figure}

\begin{figure}[bth]
\centering
\includegraphics[width=\columnwidth]{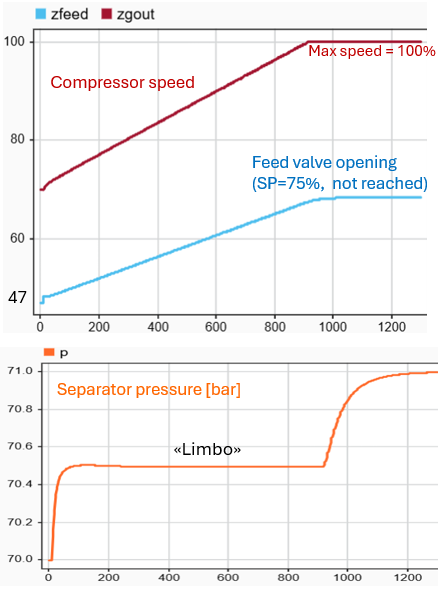}
\caption{Separator with SPC (Figure~\ref{fig:FIG3}): Dynamic simulation of setpoint increase in feed valve opening from 47\% to 75\%.}
\label{fig:SEPSIM2}
\end{figure}

\begin{figure}[bth]
\centering
\includegraphics[width=\columnwidth]{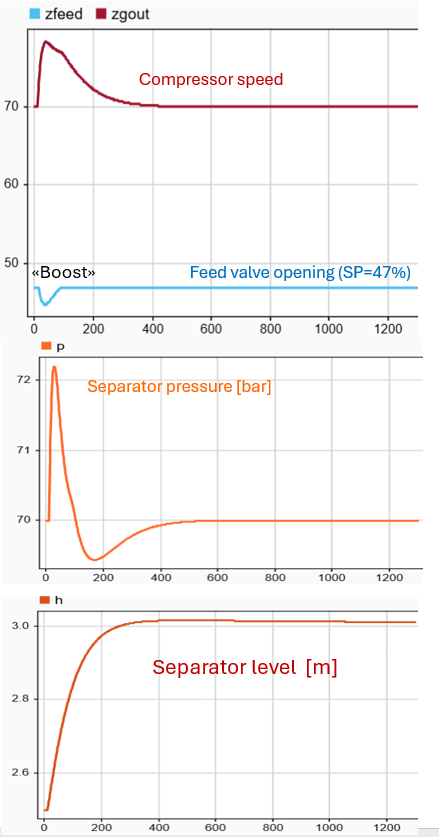}
\caption{Separator (Figure~\ref{fig:FIG3}): Dynamic simulation of setpoint increase in level ($h$) setpoint from 2.5 m to 3 m.}
\label{fig:SEPSIM3}
\end{figure}

The main objective of the simulations is to confirm that the split-parallel scheme for control of separator pressure (PCA and PCB) works as expected and to confirm that a small setpoint difference of $\Delta$=1 bar is acceptable, that is, using SPL=70 bar and SPH= 71 bar. With a small setpoint difference, one has to expect that both controllers become active in some operation regions, and as discussed by Forsman et al (JPC-paper 2026) this can be both a disadvantage (the two controllers may fight each other and we enter a ''limbo'' mode) and an advantage (the two controllers may ``boost'' the pressure response). The simple solution to avoid the limbo problem is to increase the setpoint separation $\Delta$. However, the simulations in Figure  Figures~\ref{fig:SEPSIM1} - \ref{fig:SEPSIM3} show that a small value $\Delta$=1 bar is acceptable for our case study where pressure control is fairly tight (with closed-loop time constant $\tau_c$ = 20 s for the two pressure controllers).  The controllers were tuned using the SIMC rule as explained in more detail below.

We consider a nominal operation with the feed valve setpoint at 47\% and pressure controlled using the compressor (PCA). 

Figures~\ref{fig:SEPSIM1} shows the response to a relatively small increase in the feedrate (from 47\% to 55\% valve opening). This small feedrate could be handled by the compressor alone (as illustrated in Figure~\ref{fig:SEPSIM1A}) but since the pressure setpoints are close (70 bar and 71 bar), the controller PCB is activated and keeps back the increase in the feed valve position, resulting in a ``limbo'' state with the pressure almost constant at 70.5 bar. The limbo stops when the feed valve opening reaches 55\%, so that PCB deactivates and only PCA is active and the pressure returns to 70 bar.  

In this case the limbo state of SPC, which holds back the increase in feed rate, may be an advantage as it avoids the pressure overshoot. This is seen from Figure~\ref{fig:SEPSIM1A} which shows that without controller PCB, the pressure increases much faster and goes up to almost 73 bar before returning to 70 bar. The response with conventional split range control would be the same. Thus, with split range control, the feed rate increases much faster (as given by $z_{\rm feed}$) but this comes at the expense of a large pressure overshoot. 
%, there is no limbo state and the pressure returns quickly to 70 bar, which may be an advantage. However, removing PCB has the disadvantage that pressure control is lost if the compressor saturates at 100\% speed, as illustrated next.

Figures~\ref{fig:SEPSIM2} shows the response with SPC to a large increase in the feedrate valve opening from 47\% to 75\% valve opening, which results in the compressor going to maximum speed (100\%)  (so the feed setpoint of 75\% is not quite reached). The limbo mode appears again, but this time it disappears when the compressor speed reaches its maximum of 100\%, so PCA deactivates and only PCB is active and the pressure goes to the upper setpoint of 71 bar.  

Figures~\ref{fig:SEPSIM3} shows the response to a increase in the level setpoint. In this case the activation of PCB ``boosts'' the response by making the pressure returning faster to its setpoint. 

{\bf Controller tuning}

The PI pressure controllers were tuned using the SIMC method based on obtaining first-order models using open-loop step responses. For PCA we chose $\tau_c=40$s and for PCB    $\tau_c=20$ s. For the level controller (LC) we use the $5\& 5$ rule, that is, $K_c = 5$ \%/\% (corresponding to $\tau_c = \tau/5$) and $\tau_I = 5 \tau$ where $\tau$ is the residence time based on the maximum feed rate.  The resulting PI tunings are given in Table~\ref{tab:PItunings-separator}.

\begin{table}[h]
\centering
\caption{PI tunings used in simulation of separator case study I.}
\begin{tabular}{l l l l c}
\hline
 & MV & $K_c$ & $\tau_I$ & Setpoint \\
\hline
LC &  $z_L$ & $-125\ \mathrm{\%/m}$ & $125\ \mathrm{s}$ & 2.5 m \\
PCA & $z_G$ (comp.speed) & $-2.59\ \%/\mathrm{bar}$ & $40\ \mathrm{s}$ & 70 bar \\
PCB & $z_{\rm feed}$ & $1.21\ \%/\mathrm{bar}$ & $28\ \mathrm{s}$ & 71 bar \\
\hline
\end{tabular}
\label{tab:PItunings-separator}
\end{table}

{\bf Details on the dynamic separator model}

The model is implemented in the MATLAB function
\texttt{separator1Model.m}. 
%\subsection{Variables}
The state vector is
\begin{equation}
    \boldsymbol{x}    =    \begin{bmatrix}        h \\ P    \end{bmatrix},
\end{equation}
where \(h\) [m] is the liquid level and \(P\) [bar] is the separator pressure.
The manipulated-variable vector is
\begin{equation}
    \boldsymbol{u}
    =
    \begin{bmatrix}
        z_{\mathrm{feed}} \\
        z_{\mathrm{L}} \\
        z_{\mathrm{G}}
    \end{bmatrix},
\end{equation}
where
\begin{itemize}
    \item \(z_{\mathrm{feed}}\) is the feed-valve opening (0-100\%),
    \item \(z_{\mathrm{L}}\) is the liquid-outlet-valve opening (0-100\%), 
    \item \(z_{\mathrm{G}}\) is the compressor speed (0-100\%).
\end{itemize}

The principal disturbances are
\begin{equation}
    \boldsymbol{d}
    =
    \begin{bmatrix}
        P_{\mathrm{well}} \\
        f_{\mathrm{G}} \\
        P_{\mathrm{L,out}}
    \end{bmatrix},
\end{equation}
where \(P_{\mathrm{well}}\) is the upstream well pressure,
\(f_{\mathrm{G}}\) is the gas mass fraction in the feed, and
\(P_{\mathrm{L,out}}\) is the downstream liquid pressure.

%\subsection{Feed model}

The total feed mass flow is described by a valve equation,
\begin{equation}
    \dot{m}_{\mathrm{feed}}
    =
    \dot{m}_{\mathrm{feed},0}
    \frac{z_{\mathrm{feed}}}{z_{\mathrm{feed},0}}
    \sqrt{
        \frac{
            \max(P_{\mathrm{well}}-P,0)
        }{
            P_{\mathrm{well},0}-P_0
        }
    }.
    \label{eq:separator-feed}
\end{equation}

The feed is divided into liquid and gas according to
\begin{align}
    \dot{m}_{\mathrm{L,in}}
    &=
    (1-f_{\mathrm{G}})\dot{m}_{\mathrm{feed}},
    \label{eq:separator-liquid-mass-in}
    \\
    \dot{m}_{\mathrm{G,in}}
    &=
    f_{\mathrm{G}}\dot{m}_{\mathrm{feed}}.
    \label{eq:separator-gas-mass-in}
\end{align}

The liquid volumetric feed rate is
\begin{equation}
    q_{\mathrm{L,in}}
    =
    \frac{\dot{m}_{\mathrm{L,in}}}{\rho_{\mathrm{L}}},
    \label{eq:separator-liquid-volume-in}
\end{equation}
and the gas molar feed rate is
\begin{equation}
    q_{\mathrm{G,in}}
    =
    \frac{\dot{m}_{\mathrm{G,in}}}{M_{\mathrm{W,G}}},
    \label{eq:separator-gas-molar-in}
\end{equation}
where \(\rho_{\mathrm{L}}\) is the liquid density and
\(M_{\mathrm{W,G}}\) is the gas molecular weight.

%\subsection{Separator geometry}

The separator is represented as a horizontal cylindrical vessel with diameter
\(D\), length \(L\), and radius $r=D/2$.
The total vessel volume is
\begin{equation}
    V_{\mathrm{tot}}=\pi r^2 L.
    \label{eq:separator-total-volume}
\end{equation}

For a liquid level \(h\), the cross-sectional liquid area is
\begin{equation}
    A_{\mathrm{L}}(h)
    =
    r^2
    \cos^{-1}\left(\frac{r-h}{r}\right)
    -
    (r-h)\sqrt{2rh-h^2}.
    \label{eq:separator-liquid-area}
\end{equation}

The liquid and gas volumes are therefore
\begin{align}
    V_{\mathrm{L}}(h)
    &=
    L A_{\mathrm{L}}(h),
    \label{eq:separator-liquid-volume}
    \\
    V_{\mathrm{G}}(h)
    &=
    V_{\mathrm{tot}}-V_{\mathrm{L}}(h).
    \label{eq:separator-gas-volume}
\end{align}

The derivative of the liquid volume with respect to level is
\begin{equation}
    \frac{\mathrm{d}V_{\mathrm{L}}}{\mathrm{d}h}
    =
    2L\sqrt{2rh-h^2}.
    \label{eq:separator-dvldh}
\end{equation}

In the numerical implementation, the level used in the geometrical equations
is limited according to
\begin{equation}
    h_{\mathrm{use}}
    =
    \min\left[
        \max(h,10^{-4}),D-10^{-4}
    \right],
\end{equation}
to avoid numerical problems at an empty or completely full vessel.

%\subsection{Liquid-outlet model}

The pressure immediately upstream of the liquid outlet valve includes the
hydrostatic liquid head,
\begin{equation}
    P_{\mathrm{L}}
    =
    P+\frac{\rho_{\mathrm{L}}gh}{10^5},
    \label{eq:separator-liquid-pressure}
\end{equation}
where pressures are expressed in bar.

The liquid outlet flow is described by
\begin{equation}
    q_{\mathrm{L,out}}
    =
    C_{\mathrm{v,L}}
    \frac{z_{\mathrm{L}}}{100}
    \sqrt{
        \max(P_{\mathrm{L}}-P_{\mathrm{L,out}},0)
    }.
    \label{eq:separator-liquid-out}
\end{equation}

The liquid-volume balance is
\begin{equation}
    \frac{\mathrm{d}V_{\mathrm{L}}}{\mathrm{d}t}
    =
    q_{\mathrm{L,in}}-q_{\mathrm{L,out}}.
    \label{eq:separator-liquid-balance}
\end{equation}

Combining \eqref{eq:separator-dvldh} and
\eqref{eq:separator-liquid-balance} gives the level dynamics,
\begin{equation}
    \boxed{
    \frac{\mathrm{d}h}{\mathrm{d}t}
    =
    \frac{
        q_{\mathrm{L,in}}-q_{\mathrm{L,out}}
    }{
        2L\sqrt{2rh-h^2}
    }}
    .
    \label{eq:separator-level-dynamics}
\end{equation}

The gas molar flow from the separator is assumed proportional to compressor
speed and separator pressure,
\begin{equation}
    q_{\mathrm{G,out}}
    =
    q_{\mathrm{G},0}
    \frac{z_{\mathrm{G}}}{z_{\mathrm{G},0}}
    \frac{P}{P_0}.
    \label{eq:separator-gas-out}
\end{equation}

A simplified quadratic compressor characteristic is used to calculate the
compressor discharge pressure. The compressor speed ratio is
\begin{equation}
    N_{\mathrm{r}}
    =
    \frac{z_{\mathrm{G}}}{z_{\mathrm{G},0}}.
\end{equation}

The coefficient in the quadratic compressor model is
\begin{equation}
    K_q
    =
    \frac{
        P_{\mathrm{shut},0}-P_{\mathrm{comp},0}
    }{
        q_{\mathrm{G},0}^2
    }.
    \label{eq:separator-kq}
\end{equation}

The compressor pressure rise is
\begin{equation}
    \Delta P_{\mathrm{comp}}
    =
    \max\left[
        (P_{\mathrm{shut},0}-P_0)N_{\mathrm{r}}^2
        -
        K_q q_{\mathrm{G,out}}^2,
        0
    \right].
    \label{eq:separator-compressor-rise}
\end{equation}

The compressor discharge pressure is then
\begin{equation}
    P_{\mathrm{comp}}
    =
    P+\Delta P_{\mathrm{comp}}.
    \label{eq:separator-compressor-pressure}
\end{equation}

This compressor model is used only to calculate the discharge pressure. The
gas flow in \eqref{eq:separator-gas-out} is specified independently by the
compressor speed and suction pressure.

%\subsection{Pressure dynamics}

The gas phase is described by the ideal gas law
\begin{equation}
    PV_{\mathrm{G}}
    =
    n_{\mathrm{G}}R T_{\mathrm{G}},
    \label{eq:separator-gas-law}
\end{equation}
where \(n_{\mathrm{G}}\) is the molar amount of gas in the separator,
and \(\bar{R}\) is the ideal gas constant expressed in
the appropriate units. 
The gas molar balance is
\begin{equation}
    \frac{\mathrm{d}n_{\mathrm{G}}}{\mathrm{d}t}
    =
    q_{\mathrm{G,in}}-q_{\mathrm{G,out}}.
    \label{eq:separator-gas-balance}
\end{equation}

Differentiation of \eqref{eq:separator-gas-law}, with constant temperature and
compressibility factor, gives
\begin{equation}
    V_{\mathrm{G}}\frac{\mathrm{d}P}{\mathrm{d}t}
    +
    P\frac{\mathrm{d}V_{\mathrm{G}}}{\mathrm{d}t}
    =
    Z_{\mathrm{G}}\bar{R}T_{\mathrm{G}}
    \left(
        q_{\mathrm{G,in}}-q_{\mathrm{G,out}}
    \right).
\end{equation}

Because
\begin{equation}
    \frac{\mathrm{d}V_{\mathrm{G}}}{\mathrm{d}t}
    =
    -\frac{\mathrm{d}V_{\mathrm{L}}}{\mathrm{d}t},
\end{equation}
the pressure dynamics become
\begin{equation}
    \boxed{
    \frac{\mathrm{d}P}{\mathrm{d}t}
    =
    \frac{
        Z_{\mathrm{G}}\bar{R}T_{\mathrm{G}}
        \left(q_{\mathrm{G,in}}-q_{\mathrm{G,out}}\right)
        +
        P\left(q_{\mathrm{L,in}}-q_{\mathrm{L,out}}\right)
    }{
        V_{\mathrm{G}}
    }}
    .
    \label{eq:separator-pressure-dynamics}
\end{equation}

Equations \eqref{eq:separator-level-dynamics} and
\eqref{eq:separator-pressure-dynamics} constitute the two-state nonlinear
separator model.

%\subsection{Model data}

The nominal operating conditions and model parameters are listed in
Table~\ref{tab:separator-data}.

\begin{table}[htbp]
    \centering
    \caption{Parameters for separator case study I.}
    \label{tab:separator-data}
    \begin{tabular}{llll}
        \hline
        Symbol & Description & Value & Unit \\
        \hline
        \(h_0\)
        & Nominal liquid level
        & \(2.5\)
        & \(\mathrm{m}\)
        \\

        \(P_0\)
        & Nominal separator pressure
        & \(70\)
        & \(\mathrm{bar}\)
        \\

        \(P_{\mathrm{well},0}\)
        & Nominal well pressure
        & \(180\)
        & \(\mathrm{bar}\)
        \\

        \(P_{\mathrm{L,out},0}\)
        & Nominal outlet pressure
        & \(20\)
        & \(\mathrm{bar}\)
        \\

        \(P_{\mathrm{comp},0}\)
        & Nominal discharge pressure
        & \(150\)
        & \(\mathrm{bar}\)
        \\

        \(P_{\mathrm{shut},0}\)
        & Compressor shut-off pressure
        & \(180\)
        & \(\mathrm{bar}\)
        \\

        \(z_{\mathrm{feed},0}\)
        & Nominal feed-valve opening
        & \(47\)
        & \(\%\)
        \\

        \(z_{\mathrm{L},0}\)
        & Nominal liquid-valve opening
        & \(50\)
        & \(\%\)
        \\

        \(z_{\mathrm{G},0}\)
        & Nominal compressor speed
        & \(70\)
        & \(\%\)
        \\

        \(\dot{m}_{\mathrm{feed},0}\)
        & Nominal total feed mass flow
        & \(250\)
        & \(\mathrm{kg\,s^{-1}}\)
        \\

        \(f_{\mathrm{G},0}\)
        & Nominal gas mass fraction
        & \(0.20\)
        & ---
        \\

        \(q_{\mathrm{G},0}\)
        & Nominal gas molar flow
        & \(2.5\)
        & \(\mathrm{kmol\,s^{-1}}\)
        \\

        \(\rho_{\mathrm{L}}\)
        & Liquid density
        & \(850\)
        & \(\mathrm{kg\,m^{-3}}\)
        \\

        \(M_{\mathrm{W,G}}\)
        & Gas molecular weight
        & \(20\)
        & \(\mathrm{kg\,kmol^{-1}}\)
        \\

        \(D\)
        & Separator diameter
        & \(4.0\)
        & \(\mathrm{m}\)
        \\

        \(L\)
        & Separator length
        & \(12.0\)
        & \(\mathrm{m}\)
        \\

        \(V_{\mathrm{tot}}\)
        & Total separator volume
        & \(150.8\)
        & \(\mathrm{m^3}\)
        \\

        \(T_{\mathrm{G}}\)
        & Gas temperature
        & \(400\)
        & \(\mathrm{K}\)
        \\

       % \(Z_{\mathrm{G}}\)
       % & Gas compressibility factor
       % & \(1.0\)
       % & ---
       % \\

       % \(\bar{R}\)
       % & Gas constant
       % & \(0.08314\)
       % & \(\mathrm{bar\,m^3\,kmol^{-1}\,K^{-1}}\)
       % \\

        \(g\)
        & Gravitational acceleration
        & \(9.81\)
        & \(\mathrm{m\,s^{-2}}\)
        \\

        \(C_{\mathrm{v,L}}\)
        & Liquid valve coefficient
        & \(0.0664\)
        & \(\mathrm{m^3/s,bar^{1/2}}\)
        \\

        \(K_q\)
        & Compressor char. coefficient
        & \(4.8\)
        & \(\mathrm{bar/(kmol/s)^{2}}\)
        \\
        \hline
    \end{tabular}
\end{table}

At the nominal operating point, the feed contains
\begin{align}
    \dot{m}_{\mathrm{L,in},0}
    &=
    (1-f_{\mathrm{G},0})\dot{m}_{\mathrm{feed},0}
    =
    200~\mathrm{kg\,s^{-1}},
    \\
    \dot{m}_{\mathrm{G,in},0}
    &=
    f_{\mathrm{G},0}\dot{m}_{\mathrm{feed},0}
    =
    50~\mathrm{kg\,s^{-1}}.
\end{align}

Consequently,
\begin{align}
    q_{\mathrm{L,in},0}
    &=
    \frac{200}{850}
    =
    0.2353~\mathrm{m^3/s},
    \\
    q_{\mathrm{G,in},0}
    &=
    \frac{50}{20}
    =
    2.5~\mathrm{kmol/s}.
\end{align}

The nominal gas inlet flow is therefore equal to the nominal compressor gas
flow \(q_{\mathrm{G},0}\).

For \(h_0=2.5~\mathrm{m}\), the nominal liquid and gas volumes are approximately
\begin{align}
    V_{\mathrm{L},0}
    &\approx
    99.3~\mathrm{m^3},
    \\
    V_{\mathrm{G},0}
    &\approx
    51.5~\mathrm{m^3}.
\end{align}

The nominal liquid residence time, defined as the liquid inventory divided by
the nominal liquid throughput, is approximately
\begin{equation}
    \tau_{\mathrm{L},0}
    =
    \frac{V_{\mathrm{L},0}}{q_{\mathrm{L,in},0}}
    \approx
    422~\mathrm{s}
    \approx
    7.0~\mathrm{min}.
\end{equation}

\end{document}